\documentclass[aps,prb,reprint,superscriptaddress]{revtex4-2}
\usepackage{graphicx}
\usepackage{amsmath}
\usepackage{amssymb}
\usepackage{color}
\usepackage{hyperref}
\usepackage{siunitx}
\usepackage{chemformula}

\newcommand{\C}{$^\circ$C}
\newcommand{\GZA}{GdZn$_3$As$_3$}
\newcommand{\SAC}{ScAl$_3$C$_3$}
\newcommand{\RMX}{$RM_3X_3$}

\begin{document}
	
\title{Interplay of magnetic ordering and charge transport in a distorted \SAC-type \GZA}
\author{Zhiyu Zhou}
\affiliation{Key Laboratory of Quantum Materials and Devices of Ministry of Education, School of Physics, Southeast University, Nanjing 211189, China}
\author{Xiyu Chen}
\affiliation{Key Laboratory of Quantum Materials and Devices of Ministry of Education, School of Physics, Southeast University, Nanjing 211189, China}
\author{Jia-Yi Lu}
\affiliation{School of Physics, Interdisciplinary Center for Quantum Information and State Key Laboratory of Silicon and Advanced Semiconductor Materials, Zhejiang University, Hangzhou 310058, China}
\author{Junchao Zhang}
\affiliation{Key Laboratory of Quantum Materials and Devices of Ministry of Education, School of Physics, Southeast University, Nanjing 211189, China}
\author{Xiong Luo}
\affiliation{Key Laboratory of Quantum Materials and Devices of Ministry of Education, School of Physics, Southeast University, Nanjing 211189, China}
\author{Guang-Han Cao}
\affiliation{School of Physics, Interdisciplinary Center for Quantum Information and State Key Laboratory of Silicon and Advanced Semiconductor Materials, Zhejiang University, Hangzhou 310058, China}
\affiliation{Collaborative Innovation Centre of Advanced Microstructures, Nanjing University, Nanjing 210093, China}
\author{Shuai Dong}
\affiliation{Key Laboratory of Quantum Materials and Devices of Ministry of Education, School of Physics, Southeast University, Nanjing 211189, China}
\author{Zhi-Cheng Wang}
\email{wzc@seu.edu.cn}
\affiliation{Key Laboratory of Quantum Materials and Devices of Ministry of Education, School of Physics, Southeast University, Nanjing 211189, China}
\date{\today}

\begin{abstract}
	We present the synthesis and characterization of \GZA, a previously unreported variant of the \RMX\ family ($R$ = lanthanides; $M$ = Zn, Cd; $X$ = P, As), prepared in both single-crystal and polycrystalline forms. Unlike other \RMX\ compounds that crystallize in undistorted hexagonal structures, \GZA\ adopts a distorted \SAC-type orthorhombic structure with $Cmcm$ space group. Magnetic measurements demonstrate that \GZA\ undergoes a ferromagnetic transition at the Curie temperature ($T_{\mathrm{C}}$) of 6.3~K, which is unique among known \RMX\ materials. This magnetic transition is further confirmed by specific heat and electrical resistivity measurements. \GZA\ displays metallic behavior with a pronounced resistivity peak near $T_{\mathrm{C}}$, which is strongly suppressed by magnetic fields, leading to significant negative magnetoresistance. Hall effect measurements reveal a low carrier density and a clear nonlinear anomalous Hall effect in \GZA. Furthermore, both specific heat and resistivity data suggest the presence of additional magnetic transition(s) below $T_{\mathrm{C}}$, requiring further investigation. These results demonstrate that \GZA\ possesses distinct structural, magnetic, and electronic transport properties within the \RMX\ family, establishing it as an exceptional platform for investigating competing magnetic interactions in low-carrier-density rare-earth triangular-lattice systems.
\end{abstract}

\maketitle

\section{Introduction}

In recent years, rare-earth materials with two-dimensional (2D) triangular lattices (TLs) have attracted considerable attention due to their rich physical phenomena arising from magnetic frustration~\cite{leahy2022a,song2024,avers2021,li2016a}. Notable examples include the proposed quantum spin-liquid states in Yb-based TL compounds \ch{YbMgGaO4} and \ch{NaYbCh2} (Ch = O, S, Se)~\cite{li2015a,shen2016,paddison2017,liu2018b,bordelon2019}, and the observed skyrmion state in \ch{Gd2PdSi3} resulting from the interplay between the Ruderman-Kittel-Kasuya-Yosida (RKKY) interactions and geometric frustration~\cite{kurumaji2019}. Additionally, several Eu-containing TL compounds, including \ch{EuSn2As2}, \ch{EuMg2Bi2}, \ch{EuIn2As2}, and \ch{EuCd2As2}, among others, have been identified as potential hosts for various exotic magnetic topological states~\cite{Li2019a,Li2021a,Marshall2021a,Xu2019,Zhang2020a,huaDiracSemimetalTypeIV2018,maSpinFluctuationInduced2019,xuUnconventionalTransverseTransport2021,wangColossalMagnetoresistanceMixed2021}.

Among rare-earth TL systems, the layered \RMX\ family ($R$ = lanthanides; $M$ = Zn, Cd; $X$ = P, As) stands out with well-separated TL $R^{3+}$ planes ($\sim$10~\AA)~\cite{REZn3As3_and_RECd3As3,RZn3P3_and_PrCd3P3}, which enhance quantum fluctuations in these quasi-2D magnetic systems. Previous studies have primarily focused on Ce-containing members such as \ch{CeCd3P3}, \ch{CeCd3As3}, and \ch{CeZn3P3}, where the combination of $S=1/2$ spin state, strong spin-orbit coupling, and crystal electric field effects leads to pronounced quantum phenomena~\cite{CeCd3As3,CeCd3P3,CeCd3X3,Ochiai2021}. Moreover, systematic investigations of the $R$\ch{Zn3P3} series have revealed lanthanide-dependent conductivity, long-range magnetic order, and competition between the RKKY and superexchange interactions~\cite{LnZn3P3}. Recent discoveries include the $S_{\mathrm{eff}}=1/2$ ground state in \ch{NdCd3P3} and an underlying bond instability in the $R$\ch{Cd3P3} compounds~\cite{NdCd3P3,LnCd3P3}. These findings establish the \RMX\ family as an exceptional platform for investigating geometric frustration, competing magnetic correlations, and anisotropic charge transport.

Despite the growing interest in \RMX\ materials, studies on heavy rare-earth variants remain limited, primarily due to synthesis challenges~\cite{saravanan2021a}. This work reports the successful synthesis and comprehensive characterization of \GZA, a first-time synthesized member of the \RMX\ family distinguished by its unique structural and magnetic properties. Unlike most \RMX\ compounds that crystallize in the hexagonal $P6_3/mmc$ space group~\cite{REZn3As3_and_RECd3As3,RZn3P3_and_PrCd3P3}, \GZA\ adopts a distorted \SAC-type structure with $Cmcm$ symmetry at room temperature. Remarkably, \GZA\ displays ferromagnetic (FM) ordering below $T_{\mathrm{C}} = 6.3$~K, representing the first observation of such behavior in the \RMX\ family, which we attribute to dominant RKKY interactions over antiferromagnetic (AFM) superexchange. In addition, \GZA\ exhibits metallic behavior with a pronounced resistivity peak near $T_{\mathrm{C}}$. This peak is strongly suppressed under an applied magnetic field, resulting in a large negative magnetoresistance (nMR) effect around $T_\mathrm{C}$. Further magnetic anomalies are observed at 2.5~K and 1.5~K, possibly arising from competing magnetic interactions. These distinctive characteristics establish \GZA\ as both a structural and magnetic outlier within the \RMX\ series, highlighting the potential of heavy rare-earth TL systems for hosting novel quantum states and motivating further exploration of this materials class.

\section{Methods}

\subsection{Crystal growth}

\begin{figure}
	\centering
	\includegraphics[width=\linewidth]{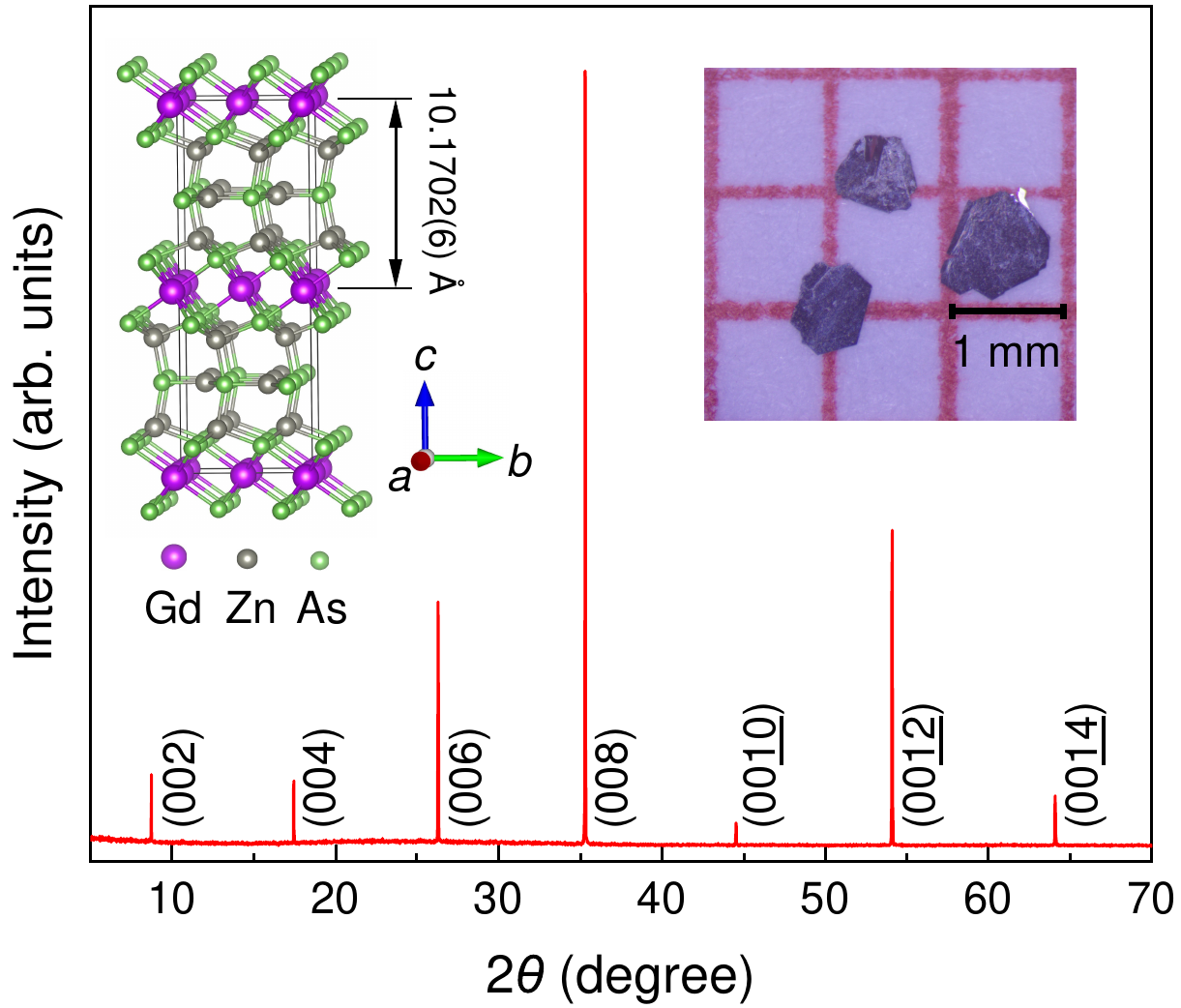}
	\caption{XRD pattern ($\theta$--2$\theta$ scan) of the \GZA\ single crystal, showing only (00$l$) reflections. The left inset displays the crystal structure, while the right inset shows a photograph of as-grown \GZA\ crystals with millimeter-scale dimensions.		
	}
	\label{fig_XRD}
\end{figure}

Single crystals of \GZA\ were synthesized using the salt-flux method. The reactants consisted of high-purity Gd ingots (99.9\%), Zn powder (99.99\%), and As lumps (99.999\%). The flux was prepared by thoroughly grinding a mixture of NaCl (99.99\%) and KCl (99.99\%) in a 1:1 molar ratio. The starting materials, Gd, Zn, and As, were mixed in a molar ratio of 1:3:3, with a total mass of 0.5 g. To prevent oxidation or hydrolysis, all reagent were handled within an argon-filled glove box. The mixture, along with approximately 4 g of flux, was sealed in evacuated double-walled silica tubes. The assembly was heated to 500 \C\ over 400 minutes, held for 24 hours, then heated to 800 \C\ over 400 minutes and held for 10,000 minutes ($\sim$7 days) before quenching in air. The thermal treatment program is sensitive; temperature deviations of just tens of kelvin or the addition of a slow cooling step may cause growth failure or impurity formation. The synthesized crystals are lustrous, brittle, millimeter-sized hexagonal platelets, as shown in the right inset of Fig.~\ref{fig_XRD}. The typical thickness of the single crystals is on the order of tens of micrometers. The x-ray diffraction (XRD) pattern of a \GZA\ single crystal, obtained using a $\theta-2\theta$ scan (Fig.~\ref{fig_XRD}), exhibits sharp (00$l$) diffraction peaks, confirming the high crystalline quality of the single crystals.

A polycrystalline (PC) \GZA\ sample was synthesized by mixing Gd, Zn, and As in a molar ratio of 1:3:3, with a total mass of 0.5 g. The resulting mixture was loaded into an alumina crucible and sealed in an evacuated silica tube. The tube was heated to 500 \C\ over 6 hours and held at that temperature for 24 hours. Subsequently, the temperature was raised to 750 \C, and the sample was maintained at this temperature for two days before free cooling. The intermediate product was then carefully ground and pressed into a pellet. The pellet was loaded into another alumina crucible, sealed in an evacuated silica tube, and heated directly to 750 \C\ over 10 hours, held at that temperature for four days, and finally allowed to cool naturally to ambient temperature. The powder x-ray diffraction (PXRD) pattern of the PC \GZA\ sample is shown in Fig.~\ref{figS1_PXRD} in the Appendix. The pattern consists of a predominant \GZA\ phase with minor impurity peaks attributed to \ch{ZnAs2}.

\subsection{Structure determination}

Suitable \GZA\ crystals were selected for single-crystal (SC) XRD analysis. Data collection was performed at 300 K using a Rigaku XtaLAB-Synergy-R diffractometer equipped with a micro-focus sealed X-ray tube (Mo $K\alpha$ radiation, $\lambda = 0.71073$ Å) and a Hypix 6000 detector. The collected frames were processed and corrected using the Rigaku OD CrysAlisPro software suite. The crystal structure was solved using the intrinsic phasing method implemented in the SHELXT structure solution program within the Olex2 environment~\cite{olex2}. Subsequently, anisotropic refinements were carried out with the SHELXL refinement package using the least-squares method~\cite{shelxle}. Finally, the crystal structure was visualized using VESTA software~\cite{vesta}. Moreover, a PANalytical x-ray diffractometer with Cu $K\alpha_1$ radiation ($\lambda = 1.5406$ Å) was used to assess the quality of both single crystals and PC samples at room temperature using $\theta-2\theta$ scans.

\subsection{Magnetization, heat capacity and resistivity measurements}

Zero-field-cooled (ZFC) and field-cooled (FC) direct-current (dc) magnetization measurements on \GZA\ were performed using a Magnetic Property Measurement System (MPMS 3, Quantum Design). Transport data were acquired using the standard four-probe technique on a Physical Property Measurement System (PPMS Dynacool, Quantum Design). In particular, resistivity data below 3 K were collected using the adiabatic demagnetization refrigerator option provided by Quantum Design and integrated into the PPMS. Heat capacity measurements were also carried out on the PPMS Dynacool using the relaxation-time method.

\section{Results and discussion}

\subsection{Crystal structure}
\begin{figure*}
    \centering
    \includegraphics[width=.8\linewidth]{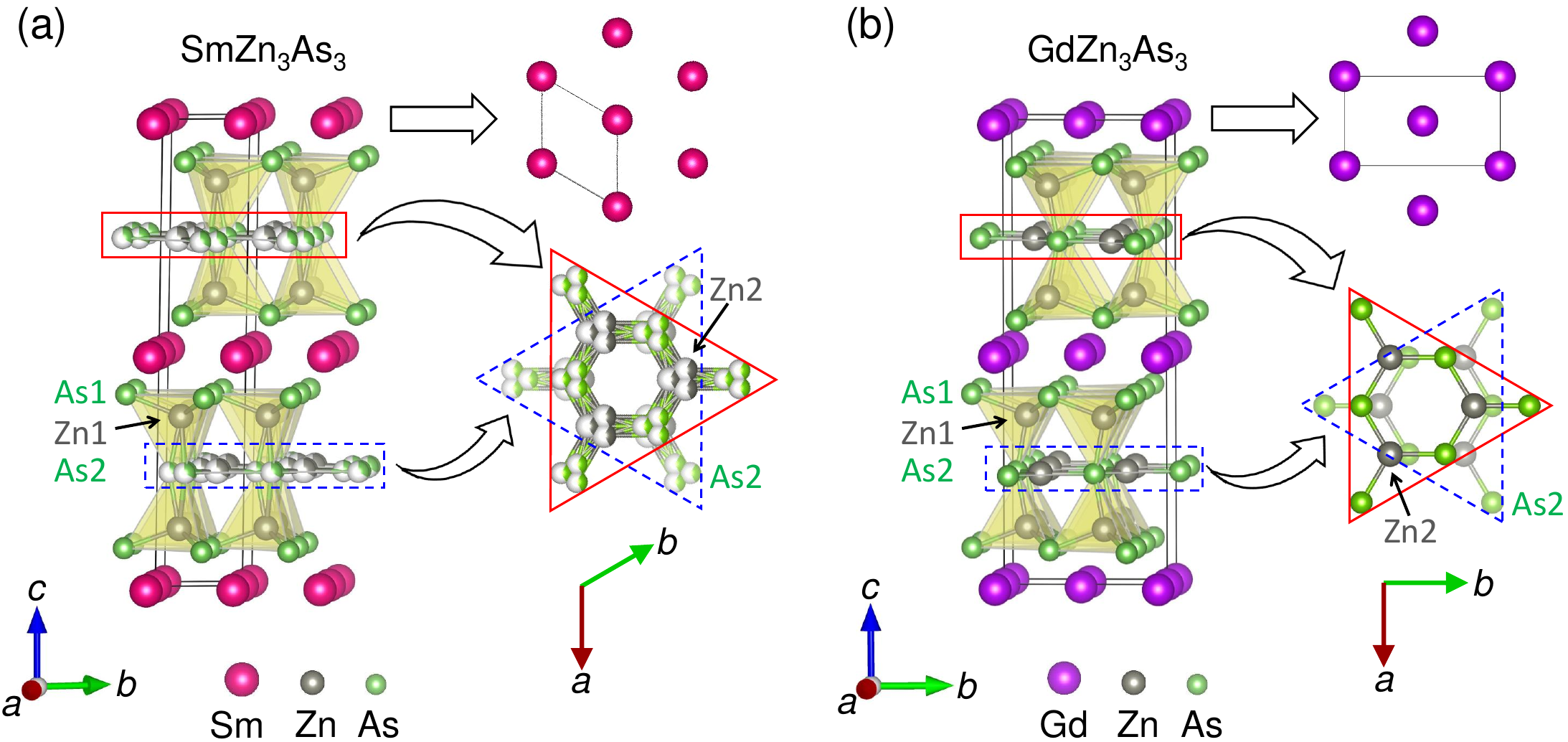}
    \caption{Comparison of the crystal structures of (a) SmZn$_3$As$_3$ ($P6_3/mmc$, hexagonal) and (b) \GZA\ ($Cmcm$, orthorhombic). The large red and purple balls represent Sm and Gd atoms, respectively. The medium grey balls correspond to Zn atoms, while the small green balls represent As atoms. The unit cells are outlined by solid black lines. The $c$-direction projections of the planar ZnAs honeycomb layers in SmZn$_3$As$_3$ and \GZA\ are displayed alongside the crystal structures, with solid red and blue dotted triangles marking two neighboring ZnAs planes.
    }
    \label{fig2_structure}
\end{figure*}

\begin{table}
\caption{\label{tab:tab1}Crystallographic parameters at room temperature and structure refinement statistics summarized for a single crystal of \GZA.}
\begin{ruledtabular}
	\begin{tabular}{lc}
	Compound & \GZA\\
	\hline
	Crystal system & Orthorhombic \\
	Space Group & $Cmcm$ (No. 63)\\
	$a$ (\AA) & 4.1311(2)\\
	$b$ (\AA) & 7.1665(3)\\
	$c$ (\AA) & 20.3403(12)\\
	$\alpha,\beta,\gamma$ ($^\circ$) & 90\\
	$V$ (\AA$^3$) & 602.19(5)\\
	$Z$ & 4\\
	$\rho_\mathrm{calc}$ (g/cm$^3$)  &  6.377 \\
	Temperature (K)  & 285.57(10)\\
	$\theta$ range (deg) & $2.0\rightarrow30.0$\\
	$\mu$ (mm$^{-1}$) & 38.885\\
	Measured reflections & 2634\\
	Independent reflections & 442\\

	$h$ & $-5\rightarrow 5$\\
	$k$ & $-8\rightarrow 9$\\
	$l$ & $-25\rightarrow 26$\\
	$R_\mathrm{int}$ & 0.0796\\
	Goodness-of-fit & 1.123\\
	$R_1\footnote{$R_1=\sum||F_o|-|F_c||/\Sigma|F_o|$.} [F^2 > 2\sigma(F^2)]$ & 0.0585\\
	$wR_2\footnote{$wR_2=\left[ \sum  w \left( F_o^2-F_c^2 \right)^2/\sum w \left( F_o^2 \right)^2 \right]^\frac{1}{2}$.} [F^2]$ & 0.1588 \\
	\end{tabular}
	\end{ruledtabular}
	
\end{table}

\begin{table*}
\caption{\label{tab:tab2} Fractional atomic coordinates, site occupancies, anisotropic displacement parameters ($U_\mathrm{ij}$), and equivalent isotropic displacement parameters ($U_\mathrm{eq}$) for each Wyckoff site in the structure of \GZA.
}
\begin{ruledtabular}
\begin{tabular}{lcccccccccc}

 Atom & Site &  $x$ & $y$ & $z$ & Occ. & $U_\mathrm{11}$~(pm$^2$)\footnote{$U_\mathrm{12} = U_\mathrm{13} = 0$} & $U_\mathrm{22}$~(pm$^2$) & $U_\mathrm{33}$~(pm$^2$) &  $U_\mathrm{23}$~(pm$^2$) & $U_\mathrm{eq}$~(pm$^2$)\footnote{$U_\mathrm{eq}$ is defined as 1/3 of of the trace of the orthogonalized $U_\mathrm{ij}$ tensor.}    \\  
\hline 
 Gd & 4$a$    & 0 & 0 & 0 & 1.00 & 1.00(8) & 1.05(8) & 1.24(9) & 0.00(5)	 & 1.10(6)      \\
 Zn1 & 8$f$    & 0 & 0.6700(3) & 0.12974(13) & 1.00 & 1.71(13) & 1.47(13) & 1.13(13) & $-$0.15(9)	 & 1.44(7) \\
 Zn2 & 4$c$    & 0 & 0.2807(7) & 0.25 & 1.00 & 2.4(2) & 2.3(2) & 6.9(4) & 0 & 3.86(14)
        \\
 As1 & 8$f$     & 0 & 0.3349(3) & 0.08469(10)& 1.00 & 0.9(1) & 1.08(11) & 1.03(11) &0.01(7) & 1.00(6)     \\
 As2 & 4$c$     & 0 & 0.6097(5) & 0.25 & 1.00 & 2.56(19) & 1.86(17) & 1.88(17)  & 0 & 2.10(9)        \\
\end{tabular}
	\end{ruledtabular}
\end{table*}


\GZA\ (space group $Cmcm$, No. 63) crystallizes in a distorted \SAC-type structure, distinct from other members of the \RMX\ series ($R$ = La--Nd, Sm; $M$ = Zn, Cd; $X$ = P, As), which adopt the undistorted hexagonal $P6_3/mmc$ space group~\cite{REZn3As3_and_RECd3As3,RZn3P3_and_PrCd3P3}. A representative compound, \ch{SmZn3As3}, is shown in Fig.~\ref{fig2_structure}(a) to illustrate the \RMX\ structure. The [\ch{Zn3As3}]$^{3-}$ slab consists of two outer layers of corner-sharing ZnAs$_4$ tetrahedra and a central mirror plane composed of heteroatomic ZnAs honeycomb layers. Moreover, the ZnAs$_4$ tetrahedra form trigonal bipyramidal units by sharing a vertex As atom from the central honeycomb layer. Notably, the Zn2 and As2 sites in the central ZnAs network are split into three positions to accommodate the unusually large in-plane displacement parameters $U_{11}$ and $U_{22}$, while the average structure retains the \SAC-type framework~\cite{REZn3As3_and_RECd3As3}. This approach is a common practice for the \RMX\ series, as well as for \ch{UAl3C3} and \ch{YbAl3C3}~\cite{REZn3As3_and_RECd3As3,RZn3P3_and_PrCd3P3,CeCd3X3,UAL3C3_AND_YBAL3C3}.

The SCXRD refinement of \GZA\ reveals an orthorhombic cell instead of an average hexagonal cell. This distortion arises from two structural features within the planar ZnAs network [Fig.~\ref{fig2_structure}(b)]: a contraction of one Zn2--As2 bond (2.358\,\AA, 2\% shorter than the other two bonds at 2.402\,\AA), and a relative displacement of neighboring honeycomb ZnAs planes by 0.7854\,\AA\ along the $b$ axis ($\Delta y = 0.1096$ in atomic coordinates). Notably, this displacement only has a slight impact on the triangular Gd$^{3+}$ lattice, as evidenced by the $b$-axis lattice constant (7.1665\,\AA) differing only 0.2\% from the ideal hexagonal value (7.1552\,\AA, $\sqrt{3}a$). The reliability of our structural analysis is supported by the following observations: (i) excellent agreement factors for the $Cmcm$ space group refinement; (ii) significant deterioration of agreement factors when forcing the hexagonal $P6_3/mmc$ model (including Zn2/As2 site splitting); (iii) absence of disorder warnings at Zn2/As2 sites in the residual electron density analysis using the program PLATON~\cite{Spek2003a}. Furthermore, \GZA\ exhibits an anomalously high $U_{33}$ displacement parameter for Zn2, a phenomenon also observed in other \RMX\ compounds and attributed to vibrational motion along the $c$ direction~\cite{REZn3As3_and_RECd3As3,RZn3P3_and_PrCd3P3,LnCd3P3}.

The distorted \SAC-type structure is rare among \RMX\ compounds at room temperature. Only \ch{CeZn3P3} has been proposed to exhibit a minor structural distortion based on its in-plane anisotropic properties, without detailed lattice parameters being reported~\cite{Ochiai2021}. Nevertheless, the structural distortion observed in \GZA\ is physically well justified for the following reasons. First of all, the site splitting phenomenon in \RMX\ compounds reflects an underlying bond instability, supported by structural phase transitions in related systems~\cite{REZn3As3_and_RECd3As3,RZn3P3_and_PrCd3P3}. For instance, low-temperature structural transitions occur in \ch{CeCd3P3} (127~K)~\cite{CeCd3P3}, \ch{CeCd3As3} (136~K)~\cite{CeCd3As3}, and \ch{YbAl3C3} (80~K)~\cite{Ochiai2007,YbAl3C3,Matsumura2008}, with \ch{YbAl3C3} confirmed to transform from a high-temperature hexagonal phase to a low-temperature orthorhombic phase. Furthermore, a very recent study revealed a local $Cmcm$ cell for $R$\ch{Cd3P3} ($R$ = La--Nd), characterized by two shorter Cd--P bonds and one longer bond in the honeycomb CdP plane, although long-range bond order remains frustrated~\cite{LnCd3P3}. Thus, the lattice instability in \RMX\ compounds and the preference for a distorted, orthorhombic cell are evident. 

Moreover, the $c/a$ values for $R$\ch{Zn3As3} ($R$ = Ce, Pr, Nd, Sm) range from 4.908 (\ch{CeZn3As3}) to 4.904 (\ch{SmZn3As3})~\cite{REZn3As3_and_RECd3As3}, whereas \GZA\ exhibits an exceptionally large $c/a$ value of 4.924 (or a $\sqrt{3}c/b$ value of 4.916). This is attributed to the relatively smaller ionic radius of Gd$^{3+}$ due to lanthanide contraction. The enhanced $c/a$ ratio in \GZA\ is equivalent to effective in-plane compression. Therefore, it is not surprising to observe an explicit structural in-plane distortion in \GZA\ at room temperature rather than in a low-temperature phase. Additionally, a high-temperature hexagonal phase is anticipated for \GZA\ above ambient conditions.

\subsection{Magnetic properties}

\begin{figure*}
    \centering
    \includegraphics[width=0.9\linewidth]{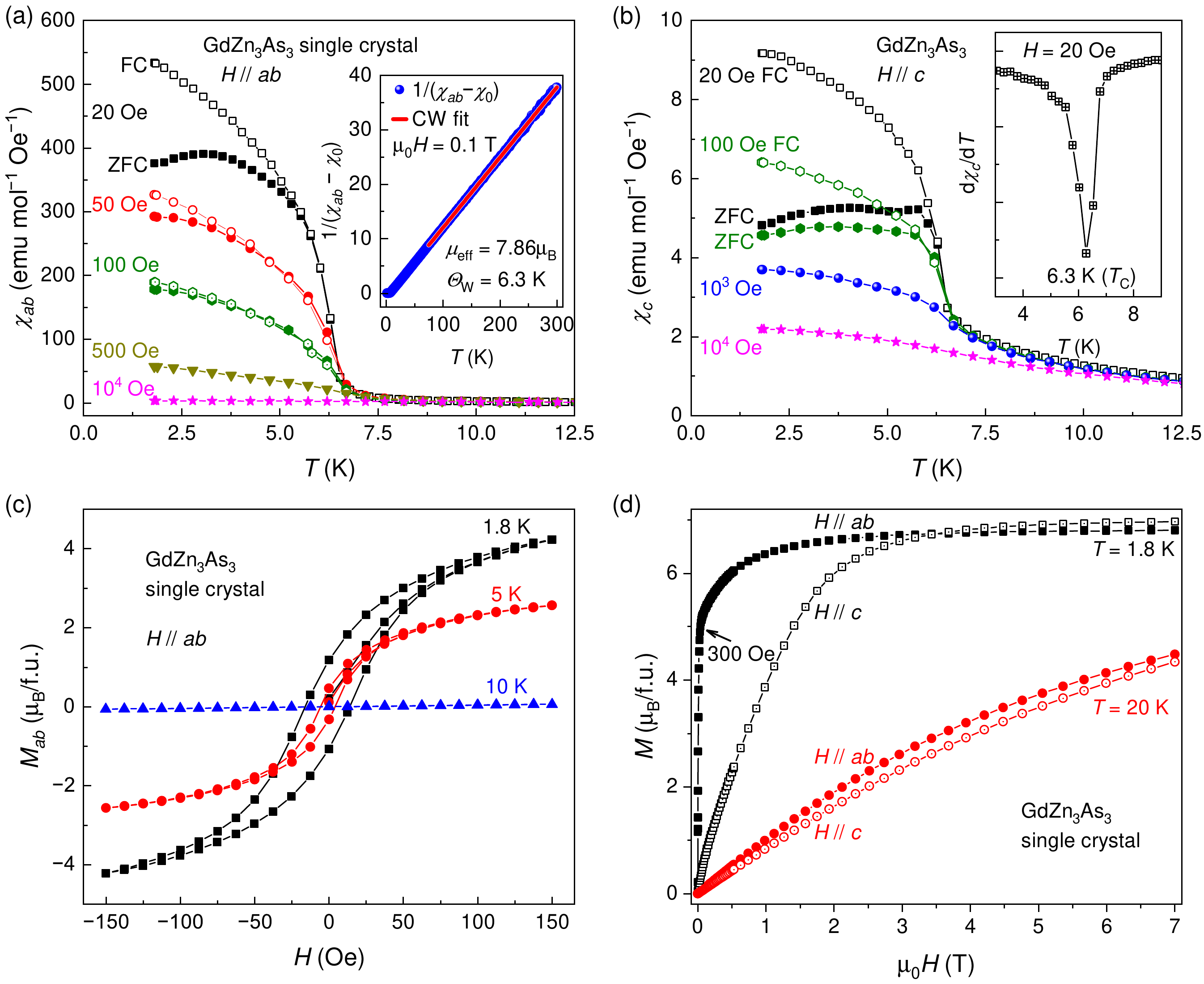}
    \caption{Magnetic properties of \GZA\ single crystals. (a) Temperature dependence of the susceptibility of \GZA\ with the magnetic field applied within the $ab$ plane. The inset shows the Curie-Weiss analysis using the data under $\mu_0H=0.1$ T, with the fitting result plotted as a red line on the inverse susceptibility. (b) Temperature-dependent susceptibility of \GZA\ with the field applied along the $c$ axis. The inset shows the temperature derivative of the susceptibility, ${d\chi}/{dT}(T)$, indicating the magnetic transition. (c) Magnetic hysteresis loops measured with in-plane fields at 1.8 K (black), 5 K (red), and 10 K (blue). (d) Magnetization curves as a function of the applied field within the $ab$ plane (solid symbols) and along the $c$ axis (dotted symbols) at 1.8 K (black) and 20 K (red).
    }
    \label{fig3_magnetism}
\end{figure*}

The magnetic properties of \GZA\ are summarized in Fig.~\ref{fig3_magnetism}. \GZA\ establishes long-range FM order with $T_\mathrm{C}$ of 6.3~K, whereas no FM order has been reported in related \RMX\ and $R$\ch{Al3C3} compounds~\cite{REZn3As3_and_RECd3As3,YbAl3C3-2,LnZn3P3,CeCd3As3,CeCd3P3}. Figures~\ref{fig3_magnetism}(a) and (b) display the temperature-dependent magnetic susceptibility ($\chi_{ab}$ and $\chi_c$) measured below 12.5~K under in-plane ($H\parallel ab$) and out-of-plane ($H\parallel c$) magnetic fields, respectively. Below 7~K, both $\chi_{ab}(T)$ and $\chi_c(T)$ exhibit a significant increase, accompanied by pronounced bifurcation between the ZFC and FC data at low fields, which are typical characteristics of a FM contribution. This bifurcation is rapidly suppressed with increasing magnetic field. The FM state of \GZA\ is further corroborated by the clear magnetic hysteresis loops in Fig.~\ref{fig3_magnetism}(c): the $M(H)$ curve at 1.8~K shows a coercive field of 30~Oe, while the hysteresis loop fully closes at 10~K, consistent with the disappearance of FM order above $T_\mathrm{C}$.

The paramagnetic $\chi_{ab}(T)$ data of \GZA\ measured at 0.1~T are analyzed via the Curie--Weiss law, as shown in the inset of Fig.~\ref{fig3_magnetism}(a). A fit of the $\chi_{ab}(T)$ data above 75~K to $\chi_{ab} = \chi_0 + C/(T - \Theta_\mathrm{W})$ yields an effective magnetic moment $\mu_\mathrm{eff} = 7.88~\mu_\mathrm{B}$ ($C = 7.76~\mathrm{emu\,K\,mol^{-1}\,Oe^{-1}}$), a Weiss temperature $\Theta_\mathrm{W} = 6.3~\mathrm{K}$, and a negligible temperature-independent term $\chi_0 = 7.32 \times 10^{-4}~\mathrm{emu\,mol^{-1}\,Oe^{-1}}$. The $\mu_\mathrm{eff}$ value closely matches the theoretical $\mu_\mathrm{eff} = 7.94~\mu_\mathrm{B}$ for Gd$^{3+}$, which is also consistent with the saturated magnetization of about 7 $\mu_\mathrm{B}$ at 1.8 K, as shown in Fig.~\ref{fig3_magnetism}(d). $\Theta_\mathrm{W}$ coincides with the FM transition temperature derived from the $d\chi_c/dT$ peak in the inset of Fig.~\ref{fig3_magnetism}(b), thus confirming intrinsic FM behavior in \GZA\ rather than canted antiferromagnetism. Additionally, PC \GZA\ samples exhibit similar magnetic hysteresis loops and an identical $T_\mathrm{C}$ value, as shown in Fig.~\ref{figS2_Pmagnetism}.

In \GZA, the TL of Gd$^{3+}$ is separated by [\ch{Zn3As3}]$^{3-}$ anionic slabs with an interlayer distance exceeding 10~\AA, suggesting weak magnetic interactions between the Gd$^{3+}$ layers. In fact, \GZA\ does exhibit pronounced magnetocrystalline anisotropy in its FM state. At 1.8~K, $\chi_{ab}$ under low fields (e.g. $\le0.01$~T) exceeds $\chi_c$ by nearly two orders of magnitude, as shown in Figs.~\ref{fig3_magnetism}(a) and (b). Figure~\ref{fig3_magnetism}(d) further demonstrates distinct anisotropic magnetization behaviors. The in-plane magnetization ($M_{ab}$) rises sharply below 300~Oe, and then gradually saturates at $\sim$0.8~T, whereas the out-of-plane magnetization ($M_c$) depends approximately linearly on the magnetic field below 1.5~T, achieving saturation at $\sim$2.5~T. The anisotropy is largely suppressed at 20~K, well above $T_\mathrm{C}$. These behaviors suggest that the Gd$^{3+}$ layer (the $ab$ plane) is the magnetic easy plane. Similar strong magnetocrystalline anisotropy also occurs in FM Eu$M_2X_2$ ($M$ = Zn, Cd; $X$ = P, As) compounds with the same $4f^7$ electronic configuration of Eu$^{2+}$~\cite{FM-EuZn2P2,FM_EuCd2P2}. For \GZA, the observed anisotropy likely related to dipole--dipole interactions, given the material's anisotropic lattice parameters and the absence of orbital contributions in the $4f^7$ state of Gd$^{3+}$~\cite{rotter2003,johnston2016}. Additionally, the presence of hole carriers in \GZA\ is revealed by magnetotransport, indicating that the FM correlations in \GZA\ are mediated by the RKKY interaction. Notably, the RKKY coupling strength may also exhibit strong crystallographic-axis dependence in low-symmetry lattices. The distorted structure of \GZA\ modifies the Fermi surface geometry, which directly governs the spatial distribution of the Fermi wavevector $k_\mathrm{F}$ --- a critical determinant of both the directional dependence and interaction-strength distribution in RKKY-mediated magnetism.

The magnetic behavior of \GZA\ stands out uniquely within the \RMX\ family. To date, FM order remains unreported in \RMX\ compounds. Most members of this family exhibit magnetic ordering below 1~K, with only two exceptions: \ch{GdZn3P3} and \ch{HoZn3P3}, which show magnetic transitions at 4.7~K (AFM) and 4.5~K (type undetermined), respectively~\cite{LnZn3P3,saravanan2021a}. The emergence of FM order in \GZA\ likely stems from the dominance of the RKKY interaction over AFM superexchange. Two key factors support this scenario. First, the larger spatial extension of As $4p$-orbitals compared to P $3p$-orbitals in \ch{GdZn3P3} enhances orbital hybridization, which may improve conductivity and strengthen FM correlations. Second, \GZA\ exhibits a resistivity one order of magnitude lower than \ch{GdZn3P3}~\cite{LnZn3P3}, suggesting higher carrier density that favors RKKY interactions over competing mechanisms.

\subsection{Heat capacity}

\begin{figure}
    \centering
    \includegraphics[width=.95\linewidth]{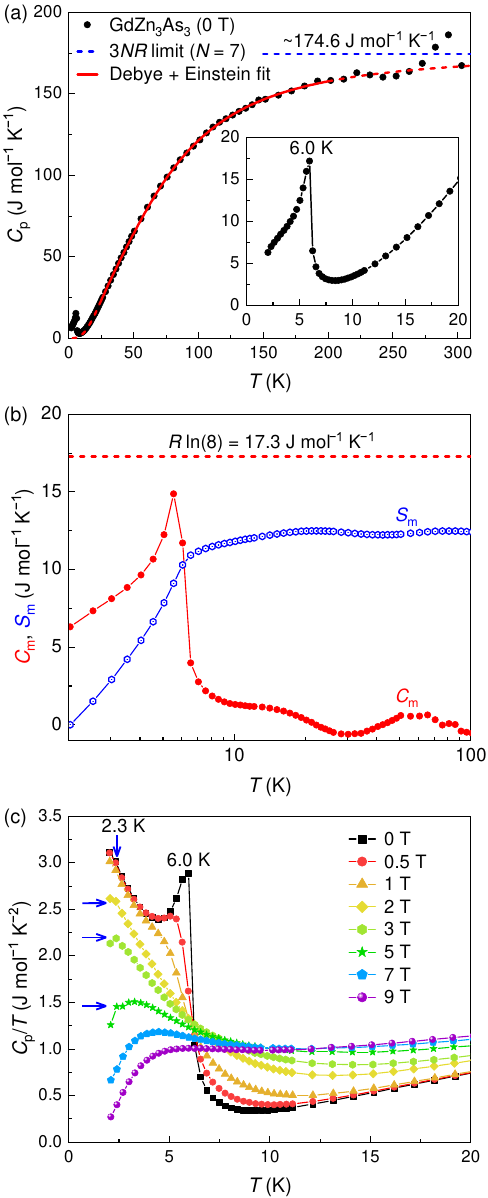}
    \caption{Heat capacity of \GZA\ single crystals. (a) Temperature dependence of the zero-field heat capacity, $C_p(T)$, fitted from 25 to 200 K using a combination of the Debye and Einstein models (red curve). The inset zooms in on the $C_p(T)$ data below 20 K, highlighting the transition at 6.0 K. (b) Magnetic heat capacity $C_\mathrm{m}(T)$ (red dots) and entropy $S_\mathrm{m}(T)$ (blue circles) derived from $C_\mathrm{m}(T)$, plotted against the logarithmically scaled temperature. The red-dashed line indicates the expected magnetic entropy for Gd$^{3+}$ ions. (c) $C_p/T$ as a function of temperature under various magnetic fields.
     }
    \label{fig4_SH}
\end{figure}

To further investigate the phase transitions in \GZA, we measured the specific heat ($C_p$), with results shown in Fig.~\ref{fig4_SH}. As shown in Fig.~\ref{fig4_SH}(a), the $C_p$ value at room temperature is slightly below the Dulong-Petit limit of $174.6~\mathrm{J\,mol^{-1}\,K^{-1}}$ theoretically calculated for \GZA. The non-smooth behavior in $C_p(T)$ near 290~K is attributed to the melting of Apiezon N grease. The lower-right inset displays a distinct $\lambda$-type peak at 6.0~K, consistent with the $T_\mathrm{C}$ of 6.3~K derived from susceptibility measurements, thereby confirming the bulk nature of the FM transition.

The zero-field $C_p$ data in the 25--200~K range were analyzed using a combination of the Debye and Einstein models according to

\begin{equation}
	C_p(T) = \alpha C_\mathrm{D}(T) + (1 - \alpha) C_\mathrm{E}(T),
	\label{eq:1}
\end{equation}
where the lattice contributions from the Debye and Einstein models are respectively expressed as
\begin{align}
	C_\mathrm{D}(T) &= 9NR \left( \frac{T}{\Theta_\mathrm{D}} \right)^3 \int_{0}^{\Theta_\mathrm{D}/T} \frac{x^4 e^x}{(e^x - 1)^2} \, dx, \label{eq:2} \\
	C_\mathrm{E}(T) &= 3NR \left( \frac{\Theta_\mathrm{E}}{T} \right)^2 \frac{e^{\Theta_\mathrm{E}/T}}{\left( e^{\Theta_\mathrm{E}/T} - 1 \right)^2}. \label{eq:3}
\end{align}
Here, $\Theta_\mathrm{D}$ and $\Theta_\mathrm{E}$ denote the Debye and Einstein temperatures, while $\alpha$ quantifies the relative contributions. The optimal fitting yields parameters $\alpha = 0.45$, $\Theta_\mathrm{D} = 142.1$~K, and $\Theta_\mathrm{E} = 287.8$~K.

Using the fitted parameters, we extracted the magnetic contribution to the heat capacity ($C_\mathrm{m}$) by subtracting the lattice component. The magnetic entropy ($S_\mathrm{m}$) was then calculated through numerical integration
\begin{equation}
	S_\mathrm{m}(T) = \int_0^T \frac{C_\mathrm{m}(T)}{T}\,dT,
	\label{eq:mag_entropy}
\end{equation}
and the results are plotted in Fig.~\ref{fig4_SH}(b). The saturated $S_\mathrm{m}$ value of $12.5~\mathrm{J\,mol^{-1}\,K^{-1}}$ reaches only 72\% of the theoretical prediction $S_\mathrm{m} = R \ln(2S + 1)$ for $S = 7/2$ ($17.3~\mathrm{J\,mol^{-1}\,K^{-1}}$). This shortfall originates from the exclusion of FM ordering contributions below 2~K, as the FM phase transition remains incomplete at this temperature boundary.

Below 30~K, $C_\mathrm{m}$ increases with decreasing temperature and tends to saturate near 15~K, forming a broad plateau. This temperature-dependent behavior of $C_\mathrm{m}$ above the magnetic ordering temperature resembles observations in \ch{CeCd3As3} and \ch{CeCd3P3}, where it was attributed to frustrated short-range AFM correlations within the triangular Ce$^{3+}$ lattices~\cite{CeCd3As3,CeCd3P3}. As previously noted, AFM superexchange interactions are prevalent in \RMX\ compounds~\cite{REZn3As3_and_RECd3As3,LnZn3P3}. Therefore, it is reasonable to propose the presence of similarly frustrated short-range AFM correlations between Gd$^{3+}$ spins above $T_\mathrm{C}$. Upon further cooling, this frustration-like behavior ultimately terminates with the establishment of a long-range FM order mediated by the RKKY interaction. The magnetic entropy $S_\mathrm{m}$ released by the frustration-like state above $T_\mathrm{C}$ is approximately $1.5~\mathrm{J\,mol^{-1}\,K^{-1}}$, accounting for only 8.7\% of the total $S_\mathrm{m}$ of Gd$^{3+}$ ($R\ln8$). This proportion is significantly smaller than those reported for \ch{CeCd3As3} and \ch{CeCd3P3}~\cite{CeCd3As3,CeCd3P3}, indicating that the magnetic behavior of \GZA\ is dominantly governed by RKKY interactions rather than superexchange interactions.

Additional anomalous features emerge when plotting $C_p/T$ versus temperature under various magnetic fields, as illustrated in Fig.~\ref{fig4_SH}(c). Below the FM transition peak, the zero-field $C_p/T$ initially decreases with lowering temperature, but then exhibits a secondary rise below 4.5~K. This upturn is attributed to another magnetic transition, since neither the electronic nor the lattice contributions to the specific heat can account for such behavior at low temperatures. Upon applying a magnetic field, an inflection point develops at 2.3~K, which intensifies progressively, eventually forming a distinct kink at 3~T and 5~T. At higher fields, this kink appears to be suppressed and merges into the broad background of magnetic heat capacity contributions. The field-dependent behavior clearly indicates that these anomalies in $C_p/T(T)$ are due to additional magnetic ordering(s) below $T_\mathrm{C}$. Multiple magnetic transitions have also been observed in \ch{SmZn3P3}~\cite{LnZn3P3}. Notably, resistivity measurements from PC \GZA\ samples reveal subtle humps near 1.5~K and 2.5~K, consistent with the anomalies observed in the specific heat data, which will be discussed in detail later.

\subsection{Magnetotransport}

\begin{figure*}
    \centering
    \includegraphics[width=\linewidth]{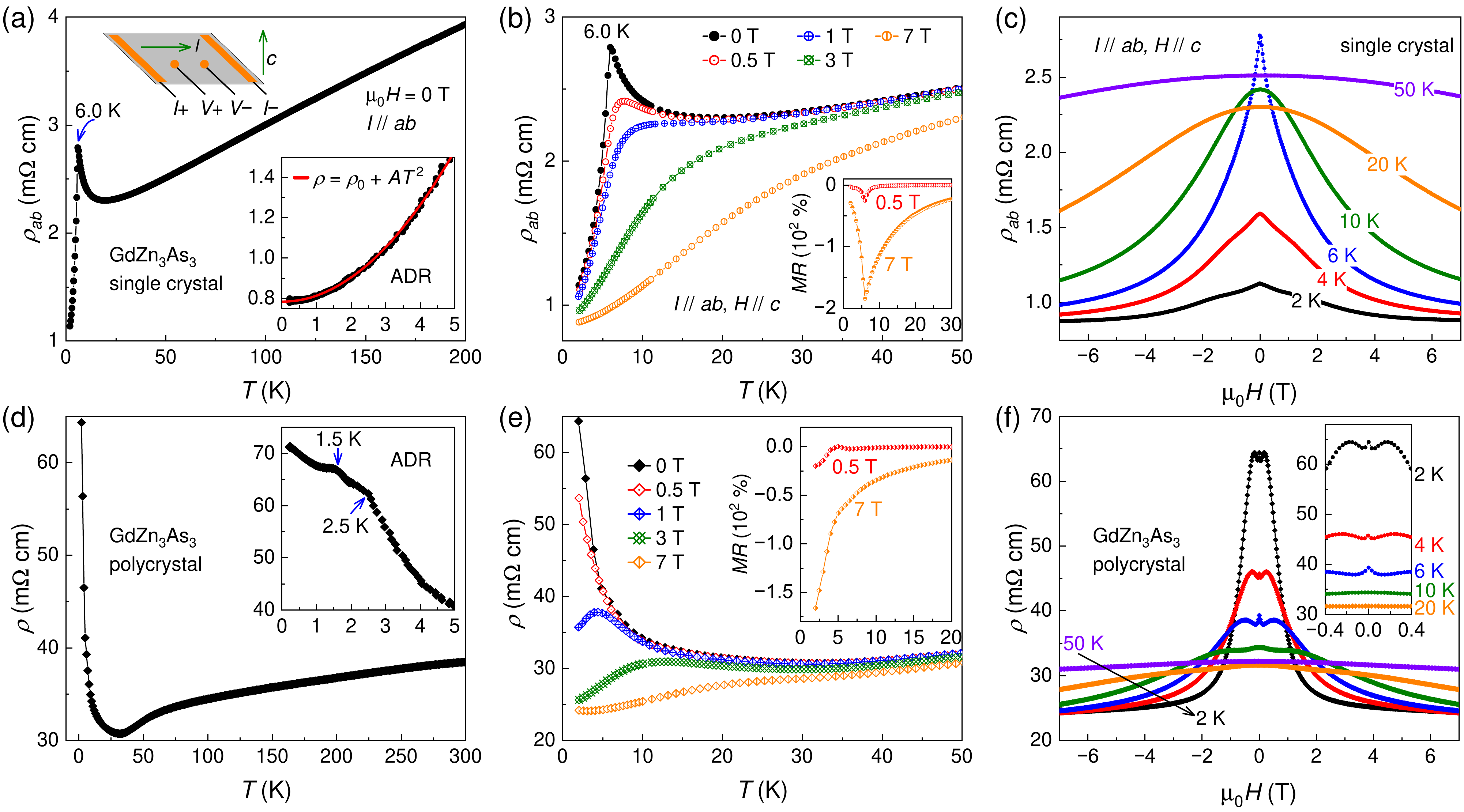}
    \caption{Comparison of the resistivity between SC (a, b, c) and PC (d, e, f) \GZA\ samples. (a, d) Temperature dependence of the electrical resistivity of \GZA\ under zero field. The upper-left inset in panel (a) shows the electrode configuration for the in-plane resistivity measurement of the SC sample. The right insets in panels (a) and (d) display the resistivity from 0.3 K to 5 K for the SC and PC samples, respectively, measured using ADR. A fit (red curve) with the parabolic function to the zero-field SC sample below 5 K is presented in the inset of panel (a). (b, e) Temperature dependence of the resistivity for the SC (b) and PC (e) samples under applied fields of 0 (black), 0.5 (red), 1 (blue), 3 (olive), and 7 T (orange). The insets in panels (b) and (e) show the calculated nMR as a function of temperature at 0.5 and 7 T. (c, f) Field dependence of the resistivity at various temperatures from 2 to 50 K for the SC (c) and PC (f) samples. The inset in panel (f) zooms in on the PC $\rho(T)$ data between $-$0.4 and 0.4 T.}
    \label{fig5_resistivity}
\end{figure*}

The electrical resistivity data of \GZA\ were collected for both SC and PC samples, as shown in Fig.~\ref{fig5_resistivity}. Panels (a)--(c) display the in-plane resistivity ($\rho_{ab}$) of the single crystal, obtained using the standard four-probe method with current applied within the $ab$ plane and magnetic fields aligned along the $c$ axis. The zero-field $\rho_{ab}(T)$ curve in Fig.~\ref{fig5_resistivity}(a) demonstrates metallic conductivity at elevated temperatures, exhibiting an approximately linear decrease with cooling. A resistivity minimum appears around 20~K, followed by a gradual upturn arising from enhanced magnetic scattering. A pronounced peak at 6.0~K marks the emergence of FM ordering, as corroborated by specific heat measurements. Below 6.0~K, the resistivity decreases sharply as magnetic scattering is suppressed by the FM alignment of spins. Remarkably, the resistivity remains below $4~\mathrm{m\Omega\,cm}$ throughout the measured temperature range, approximately one order of magnitude lower than that of \ch{GdZn3P3}~\cite{LnZn3P3}. This significant reduction is attributed to the enhanced $p$-$d$ orbital hybridization in \GZA, facilitated by the extended spatial distribution of As orbitals. A similar resistivity contrast is also observed in \ch{CeCd3As3} and \ch{CeCd3P3}~\cite{CeCd3As3,CeCd3P3}.
	
The $\rho_{ab}(T)$ data down to 0.22~K were obtained using adiabatic demagnetization refrigeration (ADR), as shown in the lower-right inset of Fig.~\ref{fig5_resistivity}(a). The data from 0 to 5~K are well described by a simple quadratic function $\rho = \rho_0 + AT^2$, yielding fitting parameters $\rho_0 = 0.784~\mathrm{m\Omega\,cm}$ and $A = 3.00 \times 10^{-2}~\mathrm{m\Omega\,cm\,K^{-2}}$. The observed $T^2$ dependence suggests that the resistivity is predominantly due to electron-electron or electron-magnon scattering processes. The characteristic Fermi liquid behavior, marked by the absence of an exponential term, indicates negligible electron-phonon scattering at low temperatures and weak electron correlations in \GZA.

Figure~\ref{fig5_resistivity}(b) displays the $\rho_{ab}(T)$ curves measured below 50~K under various magnetic fields. With increasing magnetic field, the resistivity peak systematically broadens and shifts to higher temperatures, and it essentially disappears at 7~T, resulting in significant nMR. Above 50~K, no appreciable nMR is observed for fields up to 7~T and thus these data are omitted from the figure. Using the definition $\mathrm{MR} = 100\% \times [\rho(0) - \rho(H)]/\rho(0)$, we calculate the magnetoresistance (MR) for $\rho_{ab}(T)$ curves at $\mu_0H = 0.5$~T and 7~T, with results plotted in the inset of Fig.~\ref{fig5_resistivity}(b). The nMR gradually strengthens with decreasing temperature, peaking at 6.0~K, where the 7~T curve reaches a maximum nMR of $-184\%$. Below this temperature, the nMR monotonically weakens. Figure~\ref{fig5_resistivity}(c) presents the $\rho_{ab}(H)$ curves measured at temperatures ranging from 2~K to 50~K, revealing both monotonic field dependence of $\rho_{ab}$ and maximized nMR near the transition temperature.

Owing to the brittleness and limited size of \GZA\ single crystals, coupled with the substantial contact resistance inherent to salt-flux-grown crystals, we were unable to acquire reliable measurements of the out-of-plane resistivity ($\rho_c$) in \GZA. Consequently, PC samples were synthesized for resistivity characterization to deduce $c$-direction transport properties. As demonstrated in the main panel of Fig.~\ref{fig5_resistivity}(d), the PC \GZA\ specimen exhibits resistivity behavior distinct from its SC counterpart. The room-temperature resistivity of PC \GZA\ measures approximately $40~\mathrm{m\Omega\,cm}$, with values persistently exceeding $30~\mathrm{m\Omega\,cm}$ across the entire temperature range, which is an order of magnitude greater than those observed in single crystals. Notably, the resistivity disparities between metallic SC and insulating PC forms have been documented in \ch{LaCd3P3} and \ch{CeCd3P3}, primarily attributed to minor $p$-type doping during crystal growth~\cite{CeCd3P3,CeCd3P3-PC}. However, this mechanism fails to explain the PC resistivity characteristics of \GZA. Two critical observations support this conclusion: (i) The PC \GZA\ shows metallic conduction above $\sim$30~K, in contrast to insulating behavior; (ii) Hall effect measurements (Fig.~\ref{fig6_Hall}) indicate similar carrier densities in both PC and SC \GZA. We therefore propose that the enhanced resistivity in PC \GZA\ originates from grain boundary scattering effects and the combined contributions of $\rho_{ab}$ and $\rho_c$ components. In highly anisotropic materials like \GZA, where $\rho_c \gg \rho_{ab}$ typically holds, the random crystallographic orientation in polycrystalline samples results in a weighted average dominated by the larger $\rho_c$ component.

The PC \GZA\ sample exhibits a nearly linear temperature dependence of resistivity above 100~K, resembling the behavior of its SC counterpart. Upon cooling, the resistivity shows an accelerated decrease near 60~K and reaches a minimum at 31~K. This anomalous slope change can be attributed to the superposition of $\rho_{ab}$ and $\rho_c$ components, as no phase transition is detected around 60~K. Below 30~K, a pronounced resistivity upturn emerges, which may originate from either: (i) weak localization effects induced by strong grain boundary scattering, or (ii) the enhancement of $\rho_c$ at low temperatures. Remarkably, no discernible anomaly is observed at the FM transition temperature. This discrepancy with the SC transport behavior can be rationalized by two concurrent mechanisms: First, the charge transport in PC \GZA\ is governed by $\rho_c$, whereas FM ordering primarily affects $\rho_{ab}$ (the resistivity within the magnetic easy plane), leaving $\rho_c$ essentially unaffected due to weak interplane magnetic coupling. Second, the finite grain size in PC samples restricts the FM domain dimensions, effectively decoupling the magnetic ordering from the charge transport properties. Consequently, the zero-field resistivity around $T_\mathrm{C}$ remains dominated by scattering at randomly oriented grain boundaries rather than FM ordering effects.

Moreover, ADR measurements on the PC \GZA\ reveal successive small humps around 2.5~K and 1.5~K, as highlighted in the inset of Fig.~\ref{fig5_resistivity}(d), which are invisible in the $\rho_{ab}(T)$ curves of the SC samples. These transition-like features corroborate anomalies observed in the $C_p/T$ data, specifically the secondary rise below 4.5~K and the kink at 2.3~K, as discussed earlier. We thus propose that additional magnetic transition(s) occur below $T_\mathrm{C}$, exerting a dominant influence on $\rho_c$ while leaving $\rho_{ab}$ largely unaffected. Further investigations are required to elucidate the origin of these consistent resistivity and specific heat anomalies.

The PC resistivity of \GZA\ exhibits a prominent nMR effect under magnetic field, as shown in Figs.~\ref{fig5_resistivity}(e) and (f). With increasing magnetic field, the resistivity decreases rapidly below 10~K, with the temperature-dependent MR displayed in the inset of Fig.~\ref{fig5_resistivity}(e). A maximum nMR value of $-166\%$ is observed at 2~K for the $\rho(T)$ curve under 7~T. The observation that the maximum nMR in PC \GZA\ occurs at 2~K instead of at $T_\mathrm{C}$ (as seen in SC \GZA) likely stems from the dominant $\rho_c$ contribution in the PC sample. Notably, $\rho(T)$ curves under fields above 1~T, where low-temperature upturns are suppressed, display temperature dependence similar to the $\rho_{ab}(T)$ curves in Fig.~\ref{fig5_resistivity}(b) under equivalent fields. Furthermore, the PC resistivity demonstrates nonmonotonic field dependence below 6~K, as illustrated in Fig.~\ref{fig5_resistivity}(f). For instance, the $\rho(H)$ curve at 2~K initially shows a slight decrease in the low-field regime ($\mu_0H < 0.05$~T), followed by a small increase below 0.2~T, before continuously decreasing with further field increase. This weak positive MR likely originates from enhanced spin scattering during field-induced alignment of ferromagnetic domains.

\subsection{Hall effect}

\begin{figure}
    \centering
    \includegraphics[width=\linewidth]{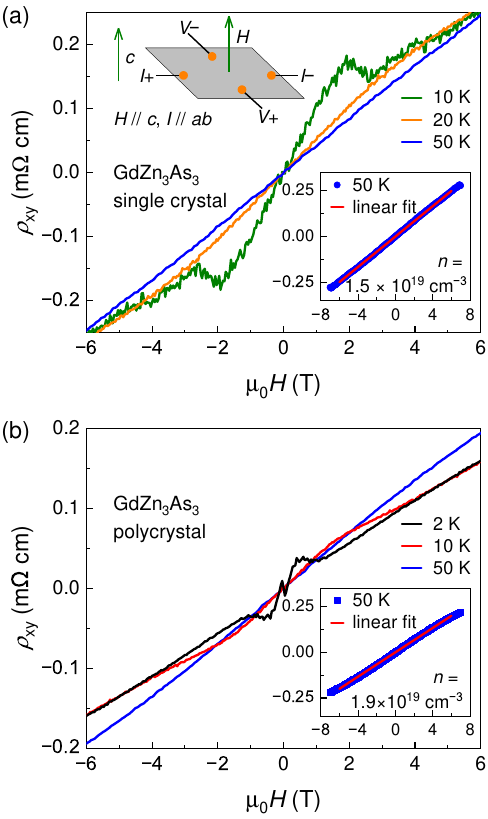}
    \caption{Magnetic field dependence of the Hall resistivity of SC (a) and PC (b) \GZA\ samples at several temperatures. The lower-right insets in both panels display the linear fits for the Hall resistivity at 50 K. For the SC \GZA, the field is applied along the $c$ axis, with the current within the $ab$ plane, as shown in the upper-left schematic of panel (a).}
    \label{fig6_Hall}
\end{figure}

Figure~\ref{fig6_Hall}(a) displays Hall effect measurements for the SC sample. Due to the large contact resistivity of salt-flux-grown single crystals, only data above 10~K exhibited acceptable quality. At 50~K, $\rho_{xy}$ shows an almost linear dependence on the magnetic field, indicating the dominance of the ordinary Hall effect (OHE). The positive slope suggests that hole-type carriers are predominant. A linear fit of of the 50~K curve yields a slope of 0.042~m$\Omega$\,cm\,T$^{-1}$, corresponding to a carrier concentration of $1.5\times 10^{19}$~cm$^{-3}$. This concentration is equivalent to $2.3\times10^{-3}$ holes per formula unit of \GZA, which is comparable to values reported for other compounds in the \RMX\ family, such as CeCd$_3X_3$ and \ch{LaCd3As3}~\cite{CeCd3As3,CeCd3P3}. As the temperature decreases, the slope of the curves exhibits minor variations, indicating a nearly constant carrier density across this range. A marked nonlinear anomalous Hall effect (AHE) is observed at 10~K, a phenomenon frequently found in numerous Eu- and Gd-containing candidate topological materials~\cite{Eu11Zn6As12,wangAnisotropyMagneticTransport2022,Giant_nonlinear_anomalous_Hall_effect,EuZn2Sb2,GdPtBi,GdPtBi-2,GdAgGe,EuCuP,EuB6}. This behavior may originate from either topological band structure or spin chirality effects within domain walls, as proposed for analogous systems~\cite{EuCd2As2-2,EuZn2Sb2,EuZn2As2}.

Hall effect measurements were also conducted on the PC \GZA\ sample, as shown in Fig.~\ref{fig6_Hall}(b). Linear fitting of the $\rho(H)$ curve at 50~K gives a slope of 0.033~m$\Omega$\,cm\,T$^{-1}$, corresponding to a hole concentration of $1.9\times 10^{19}$~cm$^{-3}$, in good agreement with the SC results. Similar to the SC case, the Hall slopes exhibit little temperature dependence. However, the PC \GZA\ displays different AHE behavior compared to the SC sample: while the SC sample shows strong nonlinear AHE at 10~K, the PC sample's nonlinear AHE reaches maximum intensity at 2~K. This distinction likely arises from the directional averaging of Hall contributions in the PC material. A clear parallel exists in EuCuP, where the anisotropic AHE manifests differently along various crystallographic directions: $\rho^A_{xy}$ ($I\parallel y$, $H\parallel z$) peaks near $T_\mathrm{C}$ (25~K), whereas both $\rho^A_{xz}$ ($I\parallel z$, $H\parallel y$) and $\rho^A_{yz}$ ($I\parallel z$, $H\parallel x$) increase monotonically with decreasing temperature, attaining maxima at 2~K~\cite{EuCuP}. This directional dependence naturally explains the different AHE characteristics observed between SC and PC \GZA\ samples.

\section{Conclusion}

To summarize, we have successfully synthesized a new variant of the \RMX\ family ($R$ = lanthanides; $M$ = Zn, Cd; $X$ = P, As), \GZA, in both SC and PC forms. Comprehensive characterization reveals an interplay between its crystal structure, magnetic order, and charge transport. Unlike most \RMX\ compounds adopting undistorted hexagonal structures, \GZA\ crystallizes in a distorted \SAC-type orthorhombic phase ($Cmcm$). While previous studies have noted structural instabilities and tendencies toward orthorhombic distortion in \RMX\ compounds, \GZA\ represents the first confirmed example that maintains this lower-symmetry structure at room temperature. This structural distortion arises from enhanced in-plane chemical pressure induced by the relatively small ionic radius of Gd$^{3+}$.

Remarkably, \GZA\ develops a strongly anisotropic FM state below $T_{\mathrm{C}}$ = 6.3~K, marking the first observation of FM ordering in the \RMX\ family. The FM correlations in \GZA\ are mediated by the RKKY interaction through low-density hole carriers. Heat capacity measurements not only confirm the bulk FM transition at $T_{\mathrm{C}}$, but also reveal additional magnetic contributions near 2.3~K, suggesting possible secondary magnetic transition(s) below the Curie temperature. Transport measurements on both SC and PC samples demonstrate metallic conductivity with pronounced nMR and comparable hole carrier concentrations. The features specific to the PC sample, including the low-temperature resistivity upturn and shifted nonlinear AHE characteristics, originate from the dominant $c$-axis transport contribution. Furthermore, ADR measurements on PC \GZA\ detect clear anomalies at 2.5~K and 1.5~K, providing additional evidence for the magnetic transitions below $T_{\mathrm{C}}$ suggested by specific heat data.

Our study reveals distinct structural, magnetic, and transport characteristics in \GZA, establishing it as a unique platform for investigating the interplay between anisotropic magnetism and charge transport. Future investigations should focus on elucidating the origin of the magnetic-ordering-like anomalies observed at 2.5~K and 1.5~K. Additionally, systematic exploration of the synthesis, crystal structure, and physical properties of $R$\ch{Zn3As3} compounds containing heavier lanthanides ($R$ = Tb--Lu) would be of significant interest for understanding the chemical trends in this material family.

\begin{acknowledgments}
	
This work was supported by the National Natural Science Foundation of China (Grants No. 12204094), the Natural Science Foundation of Jiangsu Province (Grant No. BK20220796), the Start-up Research Fund of Southeast University (Grant No. RF1028623289).
	
\end{acknowledgments}

\appendix

\section{PXRD pattern of PC \GZA}
\begin{figure*}
	\centering
	\includegraphics[width=.75\linewidth]{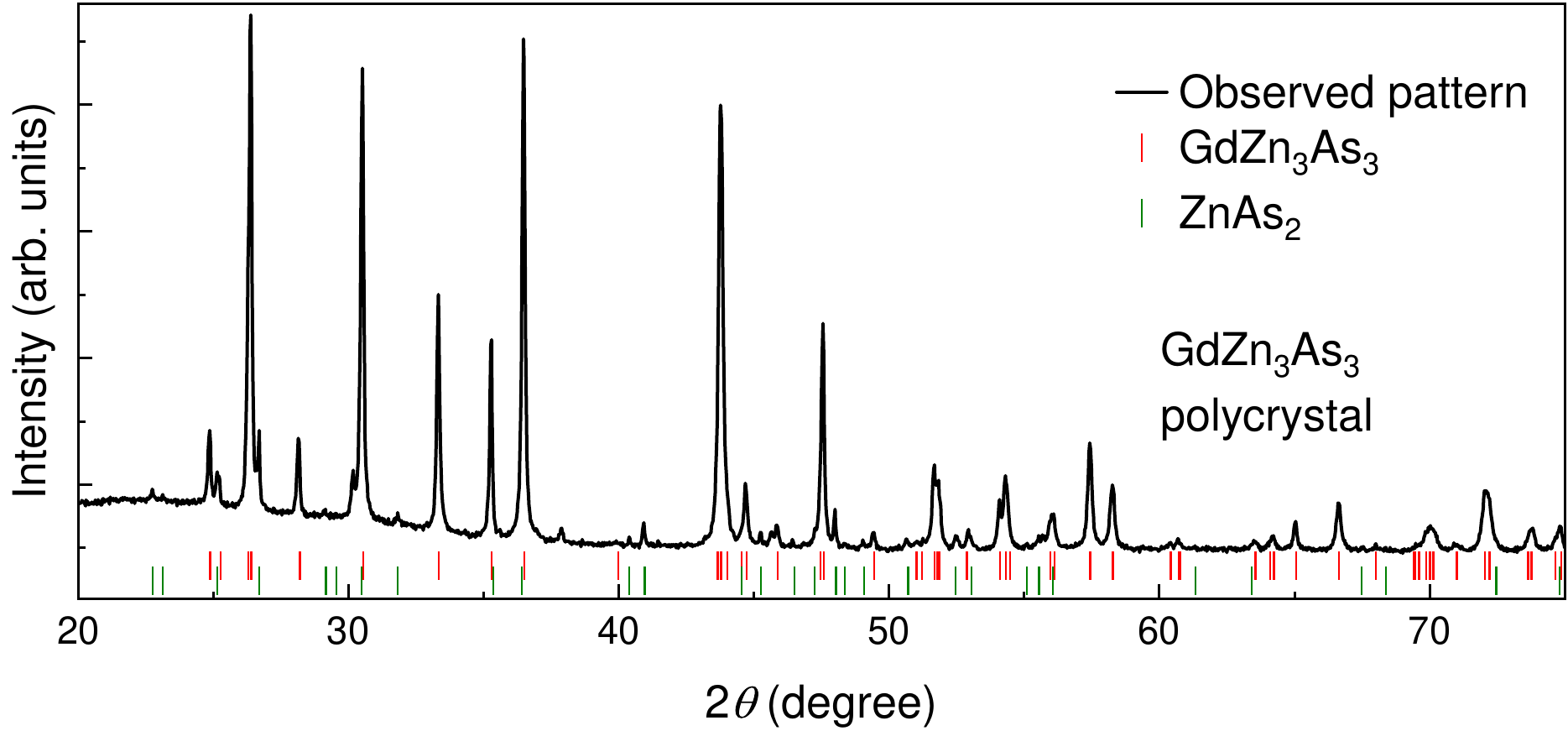}
	\caption{PXRD pattern for PC \GZA\ (black curve) and the calculated $hkl$ indices in space group $Cmcm$ (red bars). The impurity peaks originate primarily from \ch{ZnAs2} (olive bars).}
	\label{figS1_PXRD}
\end{figure*}

The PXRD pattern of PC \GZA\ is presented in Fig.~\ref{figS1_PXRD}. All major diffraction peaks can be satisfactorily indexed using the $Cmcm$ space group (indicated by red bars), confirming the successful synthesis of PC \GZA. Minor impurity peaks corresponding to \ch{ZnAs2} are identified and marked with olive bars.

\section{Magnetic properties of PC \GZA}
\begin{figure*}
	\centering
	\includegraphics[width=1\linewidth]{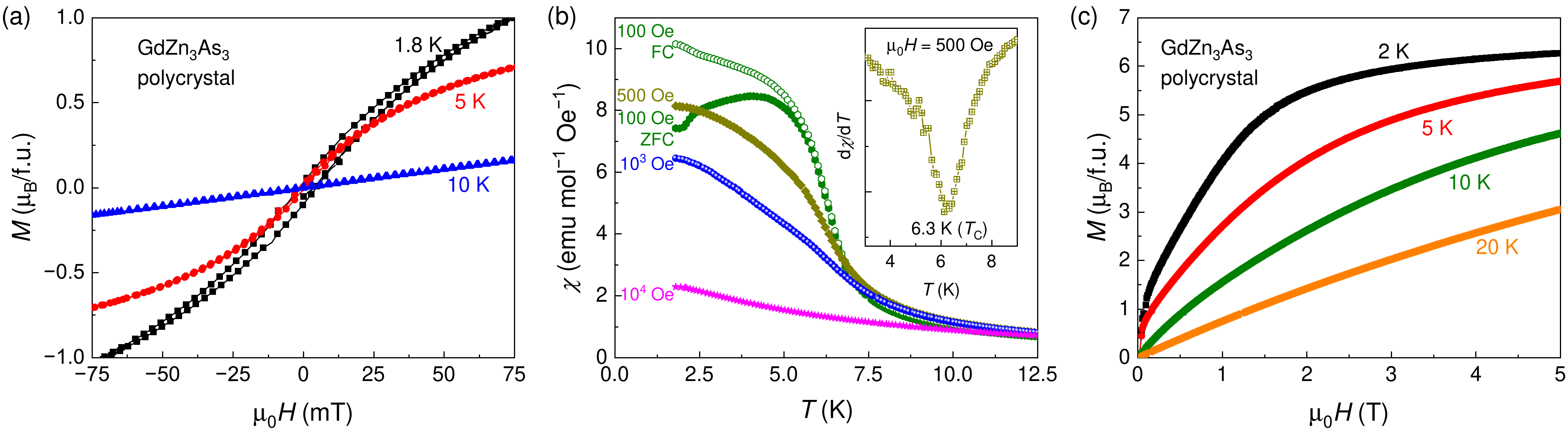}
	\caption{Magnetic properties of PC \GZA. (a) Magnetic hysteresis loops at 1.8 K (black), 5 K (red), and 10 K (blue). (b) Temperature dependence of the susceptibility under various applied fields. The inset shows ${\mathrm{d\chi}}/{\mathrm{d}T}$ versus temperature, indicating the magnetic transition at 6.3 K. (c) Magnetization as a function of the applied field at temperatures ranging from 2 to 20 K.}
	\label{figS2_Pmagnetism}
\end{figure*}

The $M(H)$ and $\chi(T)$ curves of PC \GZA\ are presented in Fig.~\ref{figS2_Pmagnetism}. While the magnetic properties of PC samples show minor discrepancies compared to SC samples due to directional signal averaging, they remain fundamentally consistent. Both the magnetic hysteresis loop at 1.8~K [Fig.~\ref{fig3_magnetism}(a)] and the pronounced ZFC/FC splitting [Fig.~\ref{fig3_magnetism}(b)] confirm the FM state previously observed in the SC sample. Furthermore, the FM transition temperature of 6.3~K, determined from the $d\chi/dT$ peak, matches exactly with the value obtained from SC susceptibility measurements.

\bibliography{reference}

\end{document}